# X-RAY EMISSION FROM JUPITER, SATURN, AND EARTH: A SHORT REVIEW

Anil Bhardwaj (*)

*NASA Marshall Space Flight Center,*
*Space Science Branch, XD12, National Space Science and Technology Center,*
*320 Sparkman Drive, Huntsville, AL 35805, USA*
*bhardwaj_spl@yahoo.com*

(*) *Now at: Space Physics Laboratory, Vikram Sarabhai Space Center,*
*Trivandrum 695022, India*
*anil_bhardwaj@vssc.org*

Jupiter, Saturn, and Earth – the three planets having dense atmosphere and a well developed magnetosphere – are known to emit X-rays. Recently, Chandra X-ray Observatory has observed X-rays from these planets, and XMM-Newton has observed them from Jupiter and Saturn. These observations have provided improved morphological, temporal, and spectral characteristics of X-rays from these planets. Both auroral and non-auroral (low-latitude) 'disk' X-ray emissions have been observed on Earth and Jupiter. X-rays have been detected from Saturn's disk, but no convincing evidence for X-ray aurora on Saturn has been observed. The non-auroral disk X-ray emissions from Jupiter, Saturn, and Earth, are mostly produced due to scattering of solar X-rays. X-ray aurora on Earth is mainly generated via bremsstrahlung from precipitating electrons and on Jupiter via charge exchange of highly-ionized energetic heavy ions precipitating into the polar atmosphere. Recent unpublished work suggests that at higher (>2 keV) energies electron bremsstrahlung also plays a role in Jupiter's X-ray aurora. This paper summarizes the recent results of X-ray observations on Jupiter, Saturn, and Earth mainly in the soft energy (~0.1-2.0 keV) band and provides a comparative overview.

## 1. Introduction

Terrestrial X-rays were discovered in the 1950s. Launch of the first X-ray satellite UHURU in 1970 marked the beginning of satellite-based X-ray astronomy. After about two decades of search with balloon-, rocket-, and satellite-based experiments[1], X-ray emission from Jupiter was discovered with the Einstein observatory[2]. During 1990s, Rontgensatellit (ROSAT) made important contributions to planetary X-rays by discovering emissions from Moon and comet and providing better observations on X-rays from Jupiter. With the advent of sophisticated X-ray observatories, viz., Chandra and XMM-Newton, the field of planetary X-ray astronomy is advancing at a faster pace. Several new solar system objects are now known to shine in X-rays at energies generally below 2 keV[3-5]. These include Venus, Mars, Saturn, Galilean moons Io and Europa, Io plasma torus, rings of Saturn, and Earth and Martian exospheres. Higher spatial and spectral resolution information on planetary X-rays is improving our understanding on the physics of the X-ray production on the planetary bodies, which are much colder than traditional million-degree K or higher temperature plasmas in the solar corona and astrophysical objects[6].

In this paper we summarize the recent results of soft (~0.1–2.0 keV) X-ray observations on Jupiter, Saturn and Earth: all the three planets having dense atmospheres and intrinsic magnetospheres, and known to emit X-rays. Table 1 provides some of the characteristic parameters of these planets. Reader are referred to other reviews for more details[1,3-5,7].

Table 1. Characteristics of Jupiter, Saturn, and Earth

| Parameter | Earth | Jupiter | Saturn |
|---|---|---|---|
| Distance from Sun (AU) | 1 | 5.2 | 9.5 |
| Equatorial Radius (Earth = 1) | 1 | 11.2 | 9.5 |
| Rotation period (Earth = 1) | 1 | 0.415 | 0.445 |
| Inclination (°) | 23.5 | 3.1 | 26.7 |
| Main atmospheric species | $N_2$, $O_2$, O | $H_2$, H, He | $H_2$, H, He |
| Magnetic field (Gauss) | 0.31 | 4.28 | 0.22 |
| Magnetic moment (Earth = 1) | 1 | 20,000 | 600 |
| Dipole tilt wrt rotation axis | +11.3° | -9.6° | -0.0° |
| Magnetosphere Size ($R_{planet}$) | 6-12 $R_E$ | 50-100 $R_J$ | 16-22 $R_S$ |
| Auroral input power (Earth=1) | 1 | $10^3$-$10^4$ | 10-100 |
| Energy Source(s) | Solar wind | rotation | Solar wind + rotation |
| Magnetospheric plasma source(s) | Ionosphere, solar wind | Io, Galilean satellites | Satellites, rings, ionosphere |

Notes: 1 AU (astronomical distance) = 1.496 x $10^{13}$ cm, Earth equatorial radius ($R_E$) = 6378 km. Average auroral input power at Earth ~ 1-100 x $10^9$ W.

## 2.  Earth: Auroral Emissions

It is well known that the X-ray aurora on Earth is generated by energetic electron bremsstrahlung[8,9,10], and the X-ray spectrum of the aurora has been very useful for studying the characteristics of energetic electron precipitation[9,11-15]. The PIXIE X-ray imager on the Polar spacecraft measured X-rays in the range 2-60 keV[16]. The high apogee of the Polar satellite (~9 $R_E$) enabled PIXIE to image the entire auroral oval with a spatial resolution of ~700 km. PIXIE data showed that the substorm X-rays brighten up in the midnight sector and have a prolonged and delayed maximum in the morning sector due to the scattering of eastward-drifting electrons[13]. Statistically, the X-ray bremsstrahlung intensity is largest in the midnight substorm onset, is significant in the morning sector, and has a minimum in the early dusk sector[10]. During the onset/expansion phase of a typical substorm the electron energy deposition power is about 60-90 GW, which produces around 10-30 MW of bremsstrahlung X-rays[17].

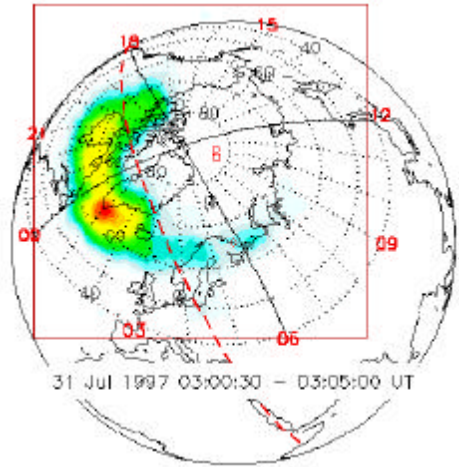

Fig 1.  Auroral X-ray image of the Earth from the Polar PIXIE instrument (energy range 2.81-9.85 keV) obtained on July 31, 1997. The red box denotes the PIXIE field-of-view. Solid black lines represent the noon-midnight and dawn-dusk magnetic local time meridians, while red dashed line represents the day/night boundary at the surface. The grid in the picture is in geomagnetic coordinates and the numbers shown in red are magnetic local time. [From PIXIE; courtesy N. Østgaard].

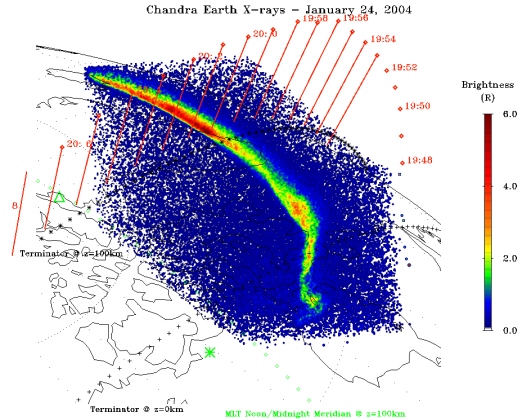

Fig. 2. Chandra HRC-I X-ray image of auroral region on January 24, 2004 showing a bright arc. The orbital location of satellite DMSP F13 is shown by red diamonds, with 2-minute time ticks and vertical lines extending down to an altitude of 100 km.

While harder X-ray emissions from electron bremsstrahlung are well known in the terrestrial aurora[9-15], surprisingly, there were no searches for emissions at auroral latitudes at energies <2 keV until recently. Northern auroral regions of Earth were imaged using the High-Resolution Camera (HRC-I) of the Chandra X-ray Observatory at 10 epochs (each ~20 min duration) between mid-December 2003 and mid-April 2004[18], to search for Earth's soft (<2 keV) X-ray aurora. The first Chandra soft X-ray observations of Earth's aurora showed that it is highly variable – sometimes intense arcs (Fig. 2), other times multiple arcs, or diffuse patches, and at times absent[18]. In at least one of the observations an isolated blob of emission is observed near the expected cusp location. A fortuitous overflight of DMSP satellite F13 provided SSJ/4 energetic particle measurements above a bright arc seen by Chandra on 24 January 2004, 20:01–20:22 UT. A model of the emissions expected strongly suggests that the observed soft X-ray signal is produced by electron bremsstrahlung[18].

## 3. Earth: Non-Auroral Disk Emissions

The non-auroral X-ray background above 2 keV from the Earth is almost completely negligible except for brief periods during major solar flares[10].

However, at energies below 2 keV soft X-rays from the sunlit Earth's atmosphere have been observed even during quite (non-flaring) Sun conditions[19,20].

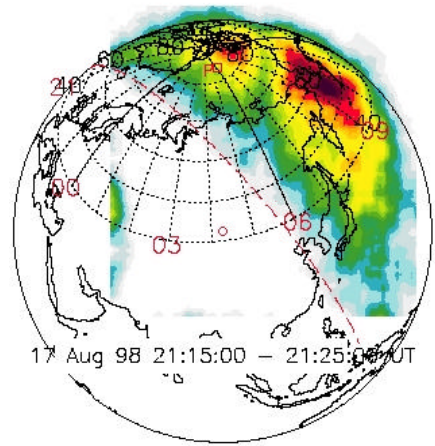

Fig. 3. X-ray image of Earth on August 17, 1998 from Polar PIXIE instrument (energy 2.9-10.1 keV), showing the dayside X-rays during a solar X-ray flare. The grid in the picture is in geomagnetic coordinates, and the numbers shown in red are magnetic local time. The day-night terminator at the surface of the Earth is shown as a red dashed line.

The two primary mechanisms for the production of X-rays from the sunlit atmosphere are: 1) the Thomson (coherent) scattering of solar X-rays from the electrons in the atomic and molecular constituents of the atmosphere, and 2) the absorption of incident solar X-rays followed by the emission of characteristic K-shell lines of Nitrogen, Oxygen, and Argon. Fig. 3 shows the PIXIE image of Earth demonstrating the X-rays (2.9–10 keV) production in the sunlit atmosphere during a solar flare of August 17, 1998. The X-ray brightness can be comparable to that of a moderate aurora. For two solar flare events during 1998 examined using the data from PIXIE, the shape of the measured X-ray spectra was in fairly good agreement with modeled spectra of solar X-rays scattered and fluoresced in the Earth's atmosphere[10].

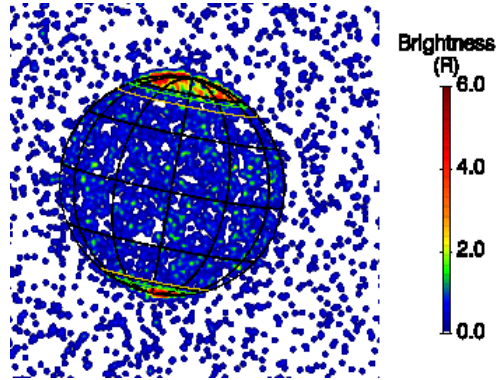

Fig. 4. Chandra X-ray image of Jupiter on 18 December 2000 generated from 10 hr of continuous observations[25]. A jovicentric graticule with $30^0$ intervals is overplotted, along with the L=5.9 (orange lines) and L=30 (green lines) footprints of the magnetic field model. The image shows strong auroral X-ray emissions from high latitudes and rather uniform emissions from the disk.

## 4. Jupiter: Auroral Emissions

Auroral X-rays from Jupiter were first detected by Einstein observatory in 1979[2], and later studied by ROSAT observations[21,22]. The pre-Chandra understanding of Jovian auroral X-rays was that these emissions are mostly line emissions resulting from recombination and charge exchange transitions in high charged states of S and O ions precipitating from inner (~8–12 $R_J$) region of the magnetosphere[1,3,21-24].

The Chandra observations of Jupiter in December 2000[25] and February 2003[26] have revealed that: 1) most of Jupiter's northern auroral X-rays come from a "hot spot" that is fixed in latitude and longitude and located significantly poleward (>30 $R_J$) of the latitudes connected to the inner magnetosphere (Figs. 4 and 8), and 2) the auroral hot spot X-rays pulsate with periodicity that is regular (~45 min) at time[25] and chaotic at other times[26] (vary over the 20–70 min range, cf. Fig. 5). Chandra observations also found (Fig. 5) that X-rays from the north and south auroral regions are neither in phase nor in anti-phase, but that the peaks in the south are shifted from those in the north by about 120º (i.e. one-third of a planetary rotation)[26]. Periodic oscillations on time scale of 20-70 min are not observed in the XMM-Newton data[27,28], perhaps due to lower spatial resolution of XMM-Newton relative to the Chandra.

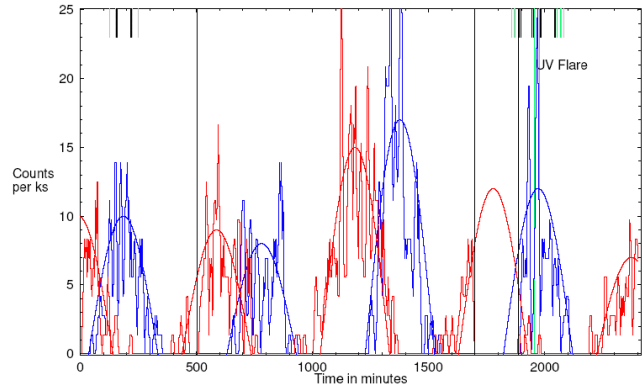

Fig. 5. Chandra Observed X ray count rate for the north (blue) and south (red) auroral zones of Jupiter[26]. The time origin corresponds to 1558:06 UT on 24 February 2003. The black vertical lines mark the transitions from ACIS-S to HRC-I exposures and back to ACIS-S. The colour bars at the top mark the simultaneous HST observations in different mode (see ref. 26 for details). Note that the set of exposures containing the UV flare coincides with the tallest peak in the ACIS-S light curve for the northern auroral zone. Smooth sections of sine waves provide crude representations of projected area effects arising from the planet's rotation.

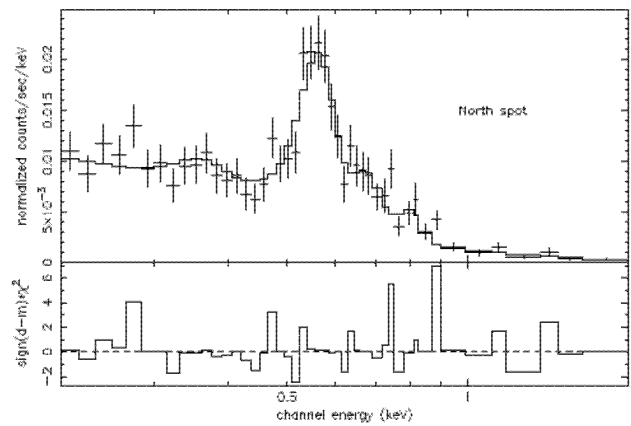

Fig. 6. XMM-Newton EPIC spectrum of Jupiter's northern auroral spot[27]. Best-fit (solid line) for the data (symbol) is obtained with five narrow Gaussian lines and a power law continuum. Bottom panel shows the chi-square of the residual. See ref. 27 for details.

The Chandra-ACIS[26] and XMM-Newton[27-29] observations have provided soft X-ray spectra from the Jovian aurora, which consist of line emissions that are consistent with high-charge states of precipitating heavy (C, O, S) ions, and not a continuum as might be expected from electron bremsstrahlung (see Figs. 6, 9a,b).

XMM-Newton has provided spectral information on the X-rays from Jupiter, which is somewhat better than Chandra. The RGS on XMM-Newton clearly resolves the strongest lines in the spectra, while the EPIC camera has provided images of the planets in the strong OVII and OVIII lines present in the Jovian auroral emissions[28,29].

The spectral interpretation of Chandra and XMM-Newton observations is consistent with a source due to energetic ion precipitation that undergoes acceleration to attain energies of >1 MeV/nucleon before impacting the Jovian upper atmosphere[26-30]. However, the source of precipitating ions – whether it's outer magnetospheric or solar wind origin, or a mixture of both, is currently not clear and arguments in favor of either of them have been presented[26-30].

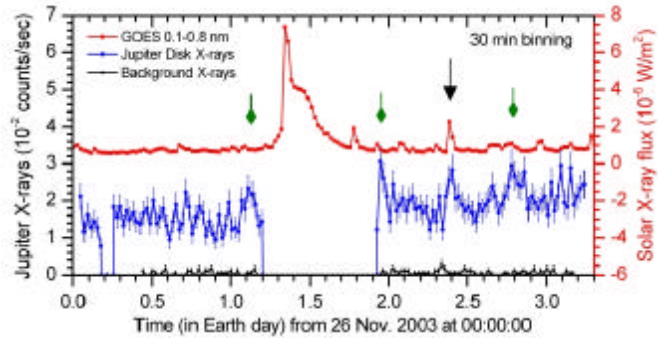

Fig. 7. Comparison of 30-min binned Jupiter disk X rays (blue curve) with GOES 10 0.1–0.8 nm solar X-ray data (red curve)[35]. The light curve of background X-rays is shown in black. The Jovian X-ray time is shifted by -4948 s to account for light travel time delay between Sun-Jupiter-Earth and Sun-Earth. The small gap at ~0.2 days is due to a loss of telemetry from XMM-Newton, and the gap between 1.2 and 1.9 days is caused by the satellite perigee passage. The black arrow (at 2.4 days) refers to the time of the largest solar flare visible from both, Earth and Jupiter, during the XMM-Newton observation, which has a clear matching peak in the Jovian light curve. The green arrows represent times when the Jupiter light curve shows peaks, which we suggest correspond to solar flares that occurred on the western (Earth-hidden) side of the Sun. The phase angle (Sun-Jupiter-Earth angle) of the observations was 10.3°, and the solar elongation (Sun-Earth-Jupiter angle) was between 76.7° and 79.8° during the observation.

Very recently, XMM-Newton and Chandra data[26] have suggested that there is a higher (>2 keV) energy component present in the spectrum of Jupiter's aurora. The observed spectrum and flux, at times, tentatively appears consistent with that produced by electron bremsstrahlung[1,32] at energies greater than 2 keV.

## 5. Jupiter: Low-Latitude Disk Emissions

X-ray emission from Jupiter's low latitudes was first reported using the ROSAT-HRI[33]. It was proposed that disk X-rays may be largely due to the precipitation of energetic sulfur or oxygen ions into the atmosphere from Jovian inner radiation belts. Later it was suggested[34] that elastic scattering of solar X-rays by atmospheric neutrals ($H_2$) and fluorescent scattering of carbon K-shell X-rays from $CH_4$ molecules located below the Jovian homopause are also potential sources of disk X-rays.

XMM-Newton's 69 hours of Jupiter observation, in Nov. 2003, demonstrated that day-to-day variation in disk X-rays of Jupiter are synchronized with variation in the solar X-ray flux (Fig. 7), including a solar flare that has a matching feature in the Jovian disk X-ray light curve[35].

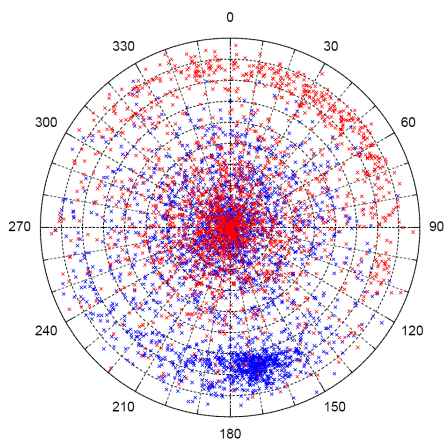

Fig. 8. Equatorial projection of X-ray photons (crosses) as seen by the Chandra ACIS-S and HRC-I instruments in the northern (blue) and southern (red) hemispheres. The concentric circles denote the latitude at intervals of 10° starting with 0° (equator) at the centre. The Jovian $S_{III}$ longitude coordinates are labeled.

The X-rays from the disk are quite uniformly distributed across the low-latitudes (Fig. 8) — in contrast to the auroral X-rays. Auroral X-rays from the north (60–75º N latitude) are dominantly confined to ~150–190º longitude and those from the south (70–80º S latitude) spread almost half-way across the planet (~300–360º and 0–120º longitude), while the disk X-rays are quite uniformly distributed and are largely confined to <50º latitude in both hemispheres[36]. The spectrum of X-rays from the disk is also harder and extends to higher energies than the auroral spectrum (Fig. 9). No periodicity has been observed in disk X-ray lightcurve[26,36,37].

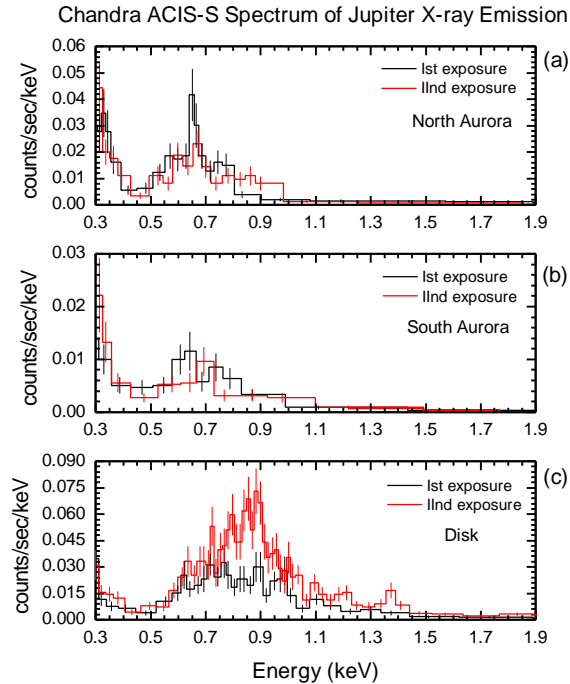

Fig. 9. Jupiter disk X-ray spectra (bottom panel, c) compared with auroral spectra from the north (top panel, a) and south (middle panel, b) observed during the two Chandra ACIS-S exposures taken on Feb. 24 and Feb. 25–26, 2003. Both ACIS-S exposures were of almost same duration and taken about a day apart. Each spectral point represents =10 events. The differences between the disk and auroral spectra are evident.

Recent studies suggested that the X-ray emission from the Jovian disk is largely due to scattered solar X-rays and that processes occurring on the Sun control the X-rays from Jupiter's disk[35,36-38].

## 6. SATURN

The X-ray emission from Saturn was unambiguously detected by XMM-Newton in October 2002[39] and by Chandra in April 2003[40]. X-rays were detected mainly from the low-latitude disk and no clear indication of auroral X-rays was observed.

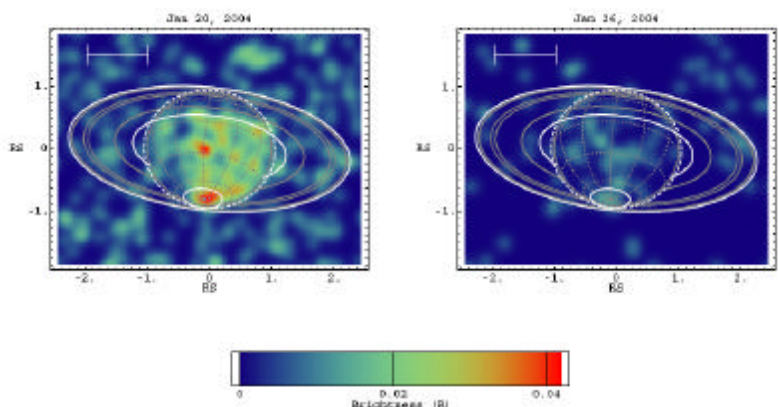

Fig. 10. Chandra ACIS X-ray 0.24-2.0 keV images of Saturn on January 20, 26, 2004[41]. Each continuous observation lasted for one full Saturn rotation. The white scale bar in the upper left of each panel represents 10?. The superposed graticule shows latitude and longitude lines at intervals of 30?. The solid gray lines are the outlines of the planet and rings, with the outer and inner edges of the ring system shown in white. The dotted white line defines the region within which events were accepted as part of Saturn's disk unless obscured by the rings. The white oval around the south pole defines the polar cap region.

Recent Observation of Saturn (Fig. 10) by Chandra in January 2004 showed that X-rays from Saturn are highly variable – a factor of 2 to 4 variability in brightness in a week's time[41]. In these observations an X-ray flare has been detected from the non-auroral disk of Saturn, which is seen in direct response to an M6-class flare emanating from a sunspot that was clearly visible from both Saturn and Earth (Fig. 11). This is the first direct evidence suggesting that Saturn's disk X-ray emission is principally controlled by processes happening on the Sun[41]. Also a good correlation has been observed between Saturn X-rays and F10.7 solar activity index. The spectrum of X-rays from Saturn disk is very similar to that from the disk of Jupiter (Fig. 12).

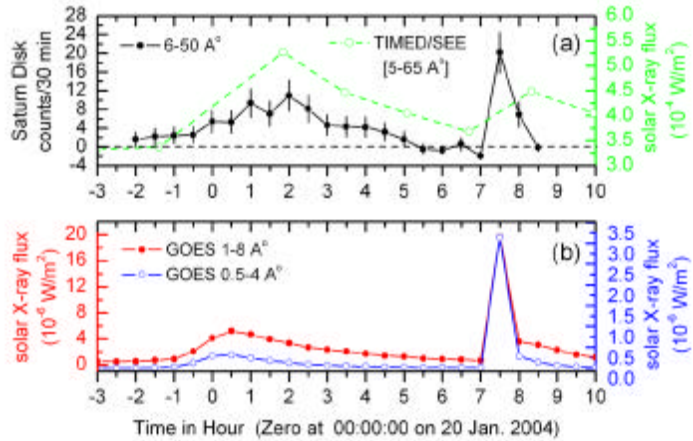

Fig. 11.  Light curve of X-rays from Saturn and the Sun on 20 January 2004[41]. All data are binned in 30 minute increments, except for the TIMED/SEE data, which are 3 minute observation-averaged fluxes obtained every orbit (~12 measurements per day). (a) Background-subtracted low-latitude (non-auroral) Saturn disk X-rays (0.24–2.0 keV) observed by Chandra ACIS, plotted in black (after shifting by -2.236 hr to account for the light-travel time difference between Sun-Saturn-Earth and Sun-Earth). The solar 0.2–2.5 keV fluxes measured by TIMED/SEE are denoted by open green circles and are joined by the green dashed line for visualization purpose. (b) Solar X-ray flux in the 1.6–12.4 and 3.1–24.8 keV bands measured by the Earth-orbiting GOES-12 satellite. A sharp peak in the light curve of Saturn's disk X-ray flux—an X-ray flare—is observed at about 7.5 hr, which corresponds in time and magnitude with an X-ray solar flare. In addition, the temporal variation in Saturn's disk X-ray flux during the time period prior to the flare is similar to that seen in the solar X-ray flux.

The Chandra observations in January 2004 also revealed X-rays from Saturn's south polar cap on Jan. 20 (see Fig. 10, left panel). However, the analysis suggests[41] that X-ray emissions from the south polar cap region on Saturn are unlikely to be auroral in nature; they might instead be an extension of its disk X-ray emission.

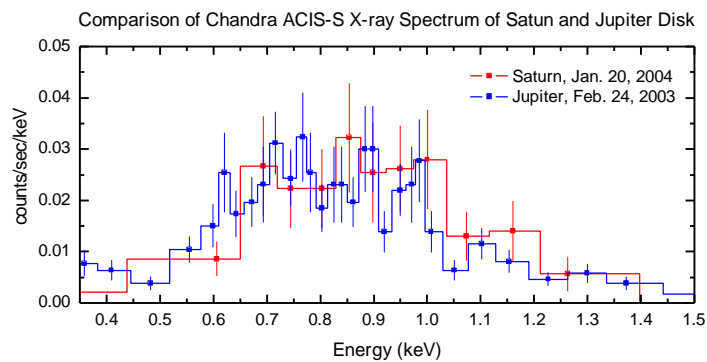

Fig. 12. Disk X-ray spectrum of Jupiter (blue curve) and Saturn (red curve). Values for Saturn spectrum are plotted after multiplying by a factor of 5.

## 7. Discussion

Table 2 presents a summary of the main characteristics of X-rays from the three planets. X-rays from the low-latitude (non-auroral) disk of all the three planets are mostly produced by scattering of solar X-rays by atmospheric species. On Jupiter and Saturn the scattering is dominantly resonant scattering with minor (~<10%) contribution from fluorescent scattering[34,38]. However, not all the incident solar X-rays in the ~0.2–2.0 keV are scattered back. The energy-average geometric X-ray albedo of Jupiter and Saturn over this energy range is ~5?10⁻⁴ [ref. 35,41]. At Jupiter precipitation of radiation belt ions can also make some contribution to the disk X-rays[33].

It has been suggested that the upper atmospheres of the giant planets Saturn and Jupiter act as "diffuse mirrors" that backscatter solar X-rays. Thus, these planets might be used as potential remote-sensing tools to monitor X-ray flaring on portions of the hemisphere of the Sun facing away from near-Earth space weather satellites[35,38,41].

The X-ray aurora on Earth is generated by energetic electron bremsstrahlung[8-10]. The auroral X-rays from Jupiter are produced by charge-exchange of highly-ionized energetic heavy ions precipitating from the outer magnetosphere and/or solar wind[1,25-30]. At higher energies (>2.0 keV) the auroral X-rays at Jupiter[31] could be produced by electron bremsstrahlung process. However, at lower (~<2.0 keV) energies electron bremsstrahlung falls short by orders of magnitude in explaining the Jupiter auroral X-ray flux. Also the spectrum shape at lower energies is inconsistent with the bremsstrahlung shape (see Figs. 6 and 9)[26,37]. At Saturn there is no clear indication of an X-ray aurora[40,41]. X-ray aurora produced by electron bremsstrahlung is expected at Saturn, but it will probably be weak and could escape detection by present-day instruments, because Saturn aurora is relatively weaker than that on Jupiter (see Table 1), and Saturn does not have copious heavy ion source, like Io on Jupiter. Recently, XMM-Newton has observed Saturn, for two planet rotations, in April and November 2005; the data is being analyzed.

Table 2.  Characteristics of X-ray emissions of Jupiter, Saturn, and Earth

| Planet | Emitting Region | Emitted Power[a] | Special Characteristics | Major Production Mechanism |
|--------|-----------------|------------------|-------------------------|----------------------------|
| Earth | Auroral atmosphere | 10-30 MW | Correlated with magnetic storm and substorm activity | Bremsstrahlung from precipitating electrons |
|  | Non-auroral atmosphere | 40 MW | Correlated with solar X-ray flux | Scattering of solar X-rays |
| Jupiter | Auroral atmosphere | 0.3-1 GW | Pulsating (~30-60 min) X-ray hot spot in north; in south emitted from a band ~180° wide in longitude | Ion precipitation (outer magnetosphere and/or solar wind) + electron bremsstrahlung |
|  | Non-auroral atmosphere | 0.3-2 GW | Relatively uniform over the disk, correlated with solar X-rays | Scattering of solar X-rays + ring current ion precipitation (?) |
| Saturn | Auroral and non-auroral atmosphere | 0.1-0.4 GW | Correlated with solar X-ray flux | Scattering of solar X-rays + Electron bremsstrahlung (?) |

[a]The values quoted are "typical" values at the time of observation. X-rays from all bodies are expected to vary with time. For comparison the total X-ray luminosity from the Sun is $10^{20}$ W.

In addition to X-rays from the planet itself, in the Jupiter system the X-ray emission has been observed from the Io plasma torus and from Galilean satellites Io and Europa[42], while in the Saturn system X-rays have been detected from the rings of Saturn[43].

## Acknowledgments

I acknowledge the fruitful collaboration with Randy Gladstone, Ron Elsner, Graziella Branduardi-Raymont, Hunter Waite, Tom Cravens, and Nikolai Østgaard on several studies reported in this paper. I also acknowledge productive collaboration with several other colleagues in the field of planetary X-rays some of whom are listed as authors or co-authors in the references mentioned in this paper. A part of this work was supported by the NRC Senior Research Associateship at the NASA Marshall Space Flight Center.